\documentclass[a4paper, amsfonts, amssymb, amsmath, reprint, showkeys, nofootinbib, twoside,longbibliography, aps]{revtex4-1}
\usepackage[english]{babel}
\usepackage[utf8]{inputenc}
\usepackage{printlen}
\usepackage[detect-all]{siunitx}
\usepackage{comment}
\usepackage[normalem]{ulem} 
\usepackage[colorinlistoftodos, color=green!40, prependcaption]{todonotes}
\usepackage{amsthm}
\usepackage{mathtools}
\usepackage{physics}
\usepackage{xcolor}
\usepackage{graphicx}
\usepackage[left=23mm,right=13mm,top=35mm,columnsep=15pt]{geometry} 
\usepackage{adjustbox}
\usepackage{placeins}
\usepackage[T1]{fontenc}
\usepackage{lipsum}
\usepackage{csquotes}
\usepackage[pdftex, pdftitle={Article}, pdfauthor={Author}]{hyperref} 

\usepackage{soul,xcolor}

\begin{document}

\title{Self-injection locking of a gain-switched laser diode}

\author{Artem E. Shitikov\textsuperscript{1}}
\email{shartev@gmail.com}
\author{Valery E. Lobanov\textsuperscript{1}}%
\author{Nikita M. Kondratiev\textsuperscript{1}}
\author{Andrey S. Voloshin\textsuperscript{2}}
\author{Evgeny A. Lonshakov\textsuperscript{1}}
\author{Igor A. Bilenko\textsuperscript{1,3}}
\affiliation{\textsuperscript{1}Russian Quantum Center, 143026 Skolkovo, Russia}
\affiliation{\textsuperscript{2}Institute of Physics, Swiss Federal Institute of Technology Lausanne (EPFL), CH-1015 Lausanne, Switzerland}
\affiliation{\textsuperscript{3}Faculty of Physics, Lomonosov Moscow State University, 119991 Moscow, Russia}

\date{April 2021}

\begin{abstract}
We experimentally observed self-injection locking regime  of the gain-switched laser to high-Q optical microresonator. We revealed that comb generated by the gain-switched laser experiences a dramatic reduce of comb teeth linewidths in this regime. We demonstrated the Lorentzian linewidth of the comb teeth of sub-kHz scale as narrow as for non-switched self-injection locked laser. Such setup allows generation of high-contrast electrically-tunable optical frequency combs with tunable comb line spacing in a wide range from 10 kHz up to 10 GHz. The characteristics of the generated combs were studied for various modulation parameters - modulation frequency and amplitude, and for parameters, defining the efficiency of the self-injection locking - locking phase, coupling efficiency, pump frequency detuning. 
\end{abstract}

\maketitle

\section{Introduction}
Narrow-linewidth lasers are in increasing demand in science and bleeding edge technologies as they give a competitive advantage in such areas as coherent communications \cite{fulop2018high}, high-precision spectroscopy \cite{Suh600,Yang:19}, optical clocks \cite{Papp:14,Newman:19},  ultrafast optical ranging \cite{Suh2018,Trocha2018,Riemensberger2020} and others. Self-injection locking (SIL) of a laser diode to a whispering gallery mode (WGM) microresonator is widely used for the linewidth narrowing and stabilization in different ranges from ultraviolet to mid-IR \cite{VASSILIEV1998305, savchenkov2019self, Liang2015, Dale:16, Savchenkov:19a, shitikov2020microresonator, Coupling18}. That technology is robust and compact. Self-injection locked laser provides a direct opportunity to generate coherent frequency combs in a form of solitonic structures \cite{ Pavlov2018, raja2019electrically, shen2020integrated, kondratiev2020numerical, lobanov2020generation, VoloshinDynamics}. Optical frequency combs may be also generated in laser diodes by rapidly switching the gain above and below the lasing threshold \cite{anandarajah2011generation}. Gain switching looks very attractive due to the simplicity of the design and it is used for spectroscopy \cite{jerez2016dual} and high-capacity communication \cite{pfeifle2015flexible}. The gain-switched (GS) lasers  also extensively use the injection locking technique for a frequency comb generation \cite{quirce2020nonlinear, zhu2016novel}, dual-comb generation \cite{quevedo2020gain}, active frequency stabilization \cite{liekhus2012injection} Kerr frequency combs excitation \cite{Wengeaba2807} and for quantum key distribution \cite{comandar2016quantum}. Injection locking can reduce the laser linewidth below 100 kHz \cite{zhou201140nm} but it is still limited by the master laser linewidth. However, self-injection locking to external passive resonator with high quality factor \cite{lin2017nonlinear}, that allows to achieve outstanding results in laser stabilization \cite{jin2020hertz, lim2017microresonator}, has never been applied to GS lasers. The implementation of the self-injection locking leads both to further miniaturization of the laser refusing the master laser and to narrowing of the laser linewidth.

In this work, we developed microresonator stabilized gain-switched laser operating in the SIL regime. We demonstrated experimentally high-contrast electrically tuned optical frequency combs with line spacing from 10 kHz to 10 GHz. It was revealed that SIL leads to a frequency distillation of each comb teeth and consequently increase the comb contrast. 
The adjustment of the modulation voltage can be used to control the width of the frequency comb in terms of the lines quantity. The widths of the central line and the comb teeth were measured as narrow as a single line in non-switched SIL regime and were equal to several kHz. That results undoubtedly will be useful for spectroscopy, multi-carrier communications, multi-frequency pump for solitonic structure generation either for anomalous or normal group velocity dispersion.

Unique combination of a gain-switched laser with self-injection locking technique allows to achieve a wide plain spectrum of the comb with an ultra-narrow sub-kHz linewidth. It leads to the bright perspective of a compact high-capacity coherent communication device construction based on ordinary easy-to-get components.

Interestingly, qualitatively comparable results of comb line distillation in the presence of the high-Q microresonator were obtained with electro-optic combs and WGM microresonator as a filter \cite{prayoonyong2021optical}. The effect of the noise reduction of the electro-optic comb lines was achieved in case of the matching of the microresonator FSR with comb line spacing which limits the range of possible modulation frequencies compare to self-injection locked GS combs.

This work consists of three parts. First, we describe the experimental setup. Second, we prove that gain-switched lasers can be self-injection locked and demonstrate the impact of the self-injection locking. Significant narrowing of the comb teeth is shown for both MHz and GHz modulation frequencies. We also study the influence of the parameters defining SIL efficiency, such as coupling efficiency, laser frequency detuning from the microresonator mode, backscattering phase, on the generated comb parameters. Third, we demonstrate the possibility of electrical tuning of comb parameters by variation of the modulation frequency and modulation depth.

\section{Experiment}

\begin{figure}[hbtp!]
\centering
\includegraphics[width=\linewidth]{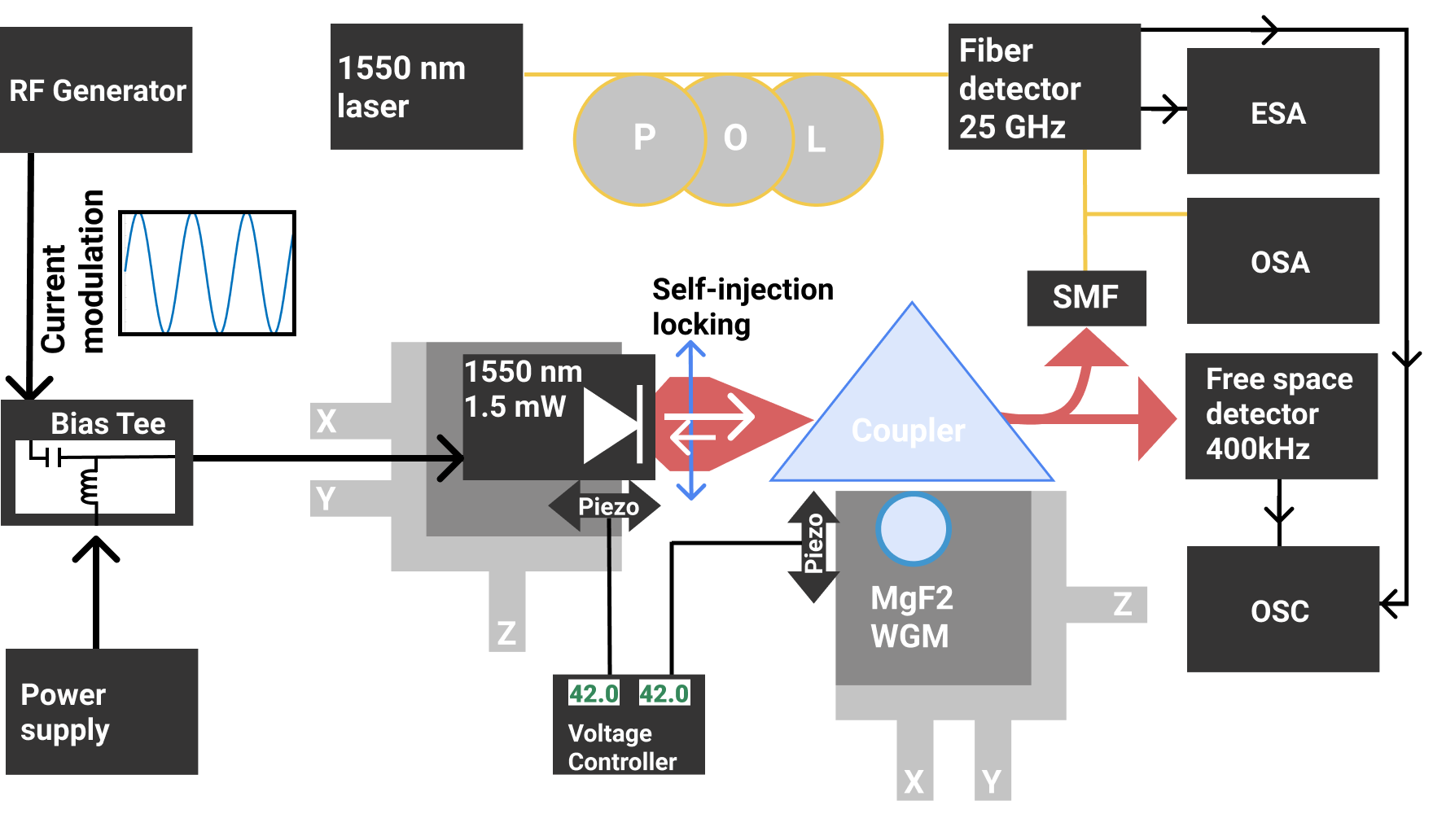}
\caption{Experimental setup. WGM modes excited with a coupling prism by DFB  1550 nm laser without isolator. Supply current was modulated with a RF generator through a bias tee. Transmitted light was analyzed on the oscilloscope, optical spectrum analyzer (OSA) and electrical spectrum analyzer (ESA). }
\label{fig:Setup}
\end{figure}

For an experimental study of the considered effect the usual SIL setup was used (see Fig. \ref{fig:Setup}). The WGMs in microresonator with diameter 4 mm made of MgF$_{2}$ were excited with a prism. A peak contrast of the mode in the transmission spectrum exceeded 30\% for the critical coupling. The internal quality factor was $4\cdot10^8$ and was controlled during experiments by the technique proposed in \cite{shitikov2020microresonator}. The 1550 nm continuous wave distributed feedback (DFB) laser without isolator was used. The locking range at a critical coupling exceeded 1 GHz. The laser power was as low as 1.5 mW which provides a linear SIL regime. The distance between the laser diode and the microresonator and between the coupler and the microresonator were precisely controlled by piezo elements. The transmitted light was split into two parts. First part was sent to the free-space detector and second one was coupled into single-mode optical fiber to observe optical spectrum or radio frequency spectrum with implementation of the heterodyne scheme.
The light coupled to a single-mode optical fiber was further mixed with a signal from narrow-linewidth fiber laser (Koheras Adjustik) and was proceeded into fiber-input detector with 40 GHz bandwidth. The signal was analyzed with oscilloscope (100 MHz bandwidth and sample rate of 2.5 GS/s) and ESA (bandwidth up to 25 GHz).

Laser diode current was modulated with a RF generator through an ordinary bias tee circuit. In our experiments we studied the impact of the modulation of the laser supply current on the SIL. Note that, when the bias tee is connected, the laser frequency shifts up and a correction of the pump current is needed to maintain the SIL regime. It can be done easily since the heterodyne laser provides us a fixed reference frequency for the real-time observation. We revealed that it is possible to stay in the SIL regime for a large range of the modulation frequencies and modulation depths. 

To prove that the laser was operated in GS regime we measured voltage from the diodes pins in case of different modulation depth. For modulation depth 8 V the minimal voltage on the pins reached 0 V during oscillations and for modulation depth 1.4 V the minimal voltage on the pins was bellow threshold of the laser, while the maximal voltage was above the threshold in both cases. The same results were obtained for modulation frequencies from 1 MHz to 100 MHz.
We also checked different waveforms as ramp and square applied to the bias tee. It didn't change the spectra dramatically, so all the results presented in this work were attained with sine modulation. 

\section{GAIN-SWITCHING IMPACT ON SIL }
We started with studying in detail how the gain switching influence on the SIL.
\begin{figure*}[htbp!]
\centering
\includegraphics[width=\linewidth]{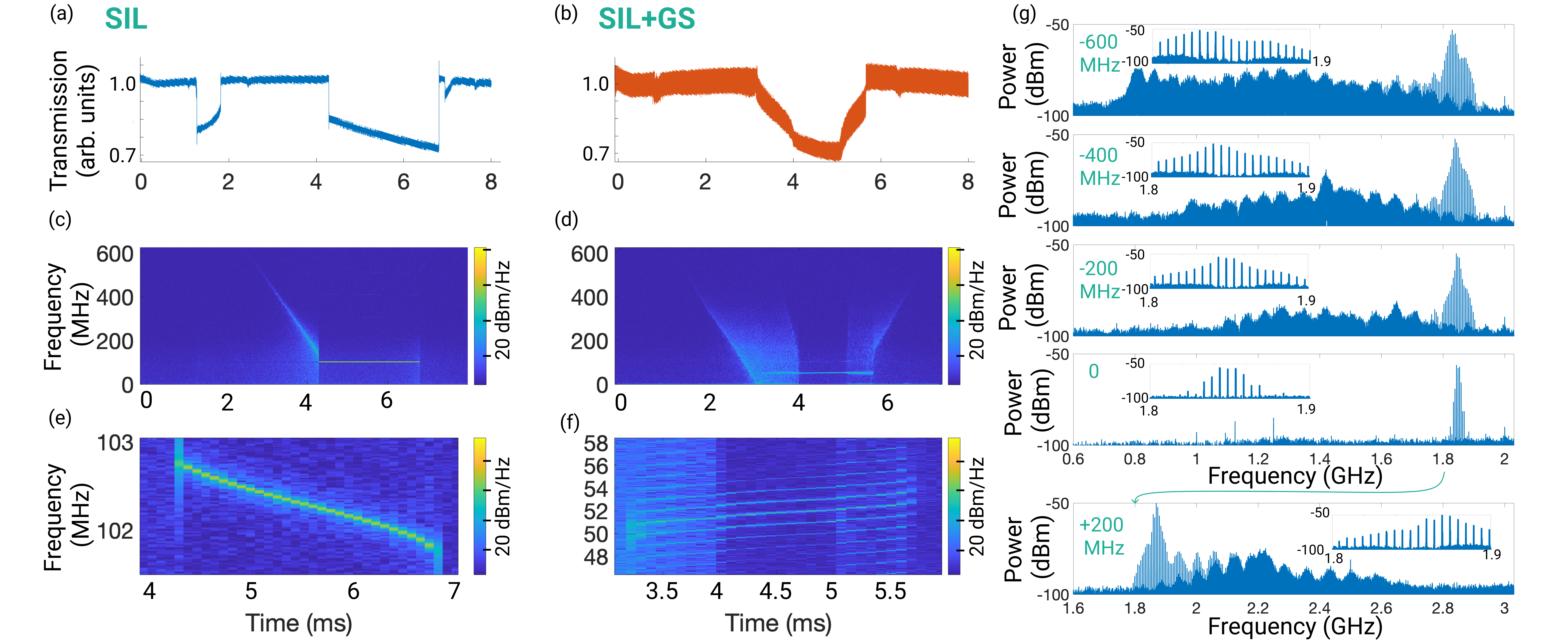}
\caption{(a)-(f) Comparison of the non-switched (left panels) and GS laser diode (right panels) in the SIL regime. (a), (b) The transmission spectra of the non-switched (a) and gain-switched (b) laser in the SIL regime. (c)-(f) The spectrograms of the (a) and (b), where (e) and (f) are enlarged area of the locking state. (g) Evolution of the output spectrum of the SIL GS laser with the variation of the detuning of the GS laser central frequency from the microresonator eigenfrequency. High-contrast lines near 1.8 GHz is the 5 MHz spaced comb (also presented in the insets), solid blue is a noisy part of the spectrum. }
\label{fig:CombsSIL}
\end{figure*}

A method of spectrograms was used for the analysis. A spectrogram is a visual representation of the frequency spectrum of a generated signal as it varies with time. Spectrograms allow to clarify the temporal dynamics of the self-injection locking. The transmitted signal from the microresonator was mixed with a signal from the reference laser and the beat-note was digitized at the oscilloscope with a sample rate up to 2.5 GS/s. We calculated short-time Fourier transform with Blackman window. The DFB laser frequency was slowly swept by the power supply and the frequency of the reference laser was constant. The transmission signal in case of  non-switched and GS SIL are presented on the (a) and (b) panels of Fig. \ref{fig:CombsSIL}. The resulting spectrograms are presented in panels (c)-(f), where (c) and (e) are for the non-switched SIL, and (d) and (f) for the gain-switched SIL with different scales. Time scales in the panels (a) and (c), as well as (b) and (d) are synchronized. The modulation frequency was equal to 1 MHz and voltage amplitude was 1.4 V. In the (c) and (d) from 2 to 4 ms one may see unlocked laser frequency evolution. Without the gain switching one blurred line is observed, and with gain switching it is changed to a broad frequency range. In the (e) and (f) the locking area is presented enlarged. In the spectrogram (f) separate comb lines can be seen. In panel (c) one may see unlocked laser frequency changing from 2.5 s to 4 s, then it becomes self-injection locked to WGM. The frequency change almost ceases in SIL regime. The ratio of the slope of the tuning curve in locked and unlocked regimes determines the stabilization coefficient. There is no unlocked part after SIL in the panel because laser frequency detuned out of the sample rate bandwidth. If there were no SIL, one would have observed V-shaped beatnote frequency dependence. In the panel (d) one may see a spectrum of the unlocked gain-switched laser from 1.5 to 3 s. Then it becomes partially locked by WGM from 3.5 to 4 s. The examples of the spectra in that case are presented in the first 3 panels from (g). Such regime corresponds to the locking of the generated sideband to the WGM. For higher modulation frequencies it leads to multiple excitations of WGM with a maximum transmission dip near the center. Then the laser becomes fully self-injection locked in interval from 4 to 5 s (4th panel in (g)). It happens when central frequency is locked to the WGM. In the interval from 5 to 5.5 s it anew transfers into a  partially locked state (last panel in (g)) and then finally one may observe a fringe of the unlocked gain-switched spectrum from 5.5 till 7 s. It is worth noting that the inclination of the lines in GS regime, which is determined by the stabilization coefficient, is close to the non-switched SIL one. In the panel (g) one can see the spectra obtained for different GS laser frequency detuning from the WGM resonance. Gain switching completely alters the SIL dynamic. As the spectrum of the GS laser approaches the SIL range, high narrow comb lines in this range appears. To achieve an optimal regime with fully distilled output spectrum one should set the detuning of the laser from the resonance close enough to zero.

\begin{figure}[hbtp!]
\centering
\includegraphics[width=0.9\linewidth]{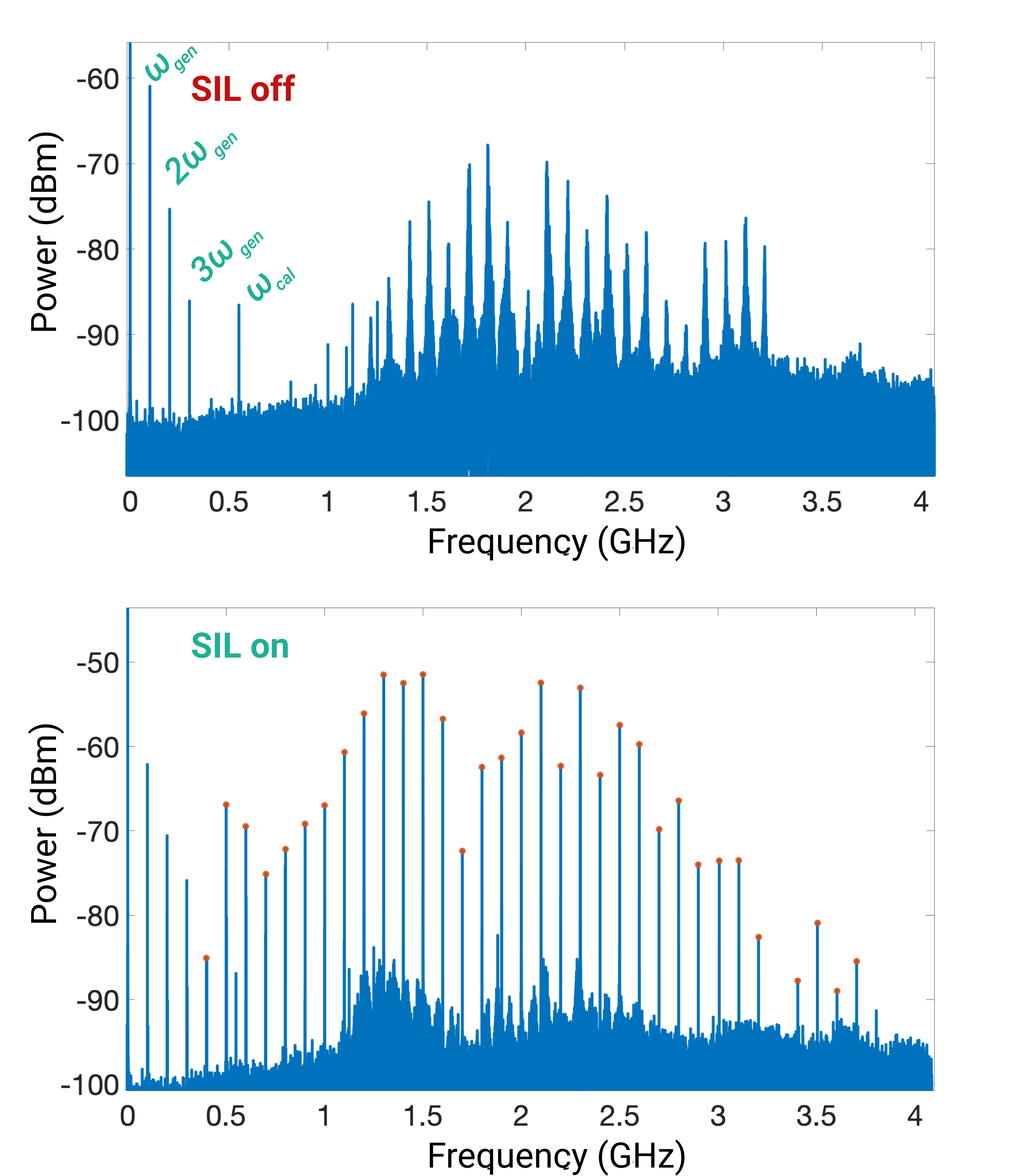}
\caption{Spectra of the 100-MHz gain-switched comb generated by the free-running (top panel) and self-injection locked (bottom panel) laser. SIL leads to noise suppression and higher power level. Lower spectrum is noticeably spectrally purified.  
}
\label{fig:OnOff}
\end{figure}

As it was shown in several works \cite{Kondratiev:17, Galiev20}, the value of the locking phase is an important parameter for SIL. We were able to tune it varying the distance between the laser and the microresonator. We found that the optimal phase for the unmodulated and modulated cases may not coincide. First, to obtain the optimal phase in the modulated case we adjust the distance between the laser and the microresonator to achieve maximum peak contrast on the free-space detector. The laser frequency was continuously scanned during the alignment process. Then the frequency scan was turned off and laser detuning was adjusted for optimal GS-comb excitation. 

Obtained results confirm that pump modulation does not lead to the suppression of the SIL regime. Interestingly and even surprisingly, the locking persists even for the modulation frequencies larger than the locking range width. This can be viewed as follows. First the self-injection locking narrows the laser linewidth, then the parametric process in the laser provides comb teeth, which are also narrow because of the narrow seed.

\begin{figure*}[htbp!]
\centering
\includegraphics[width=0.9\linewidth]{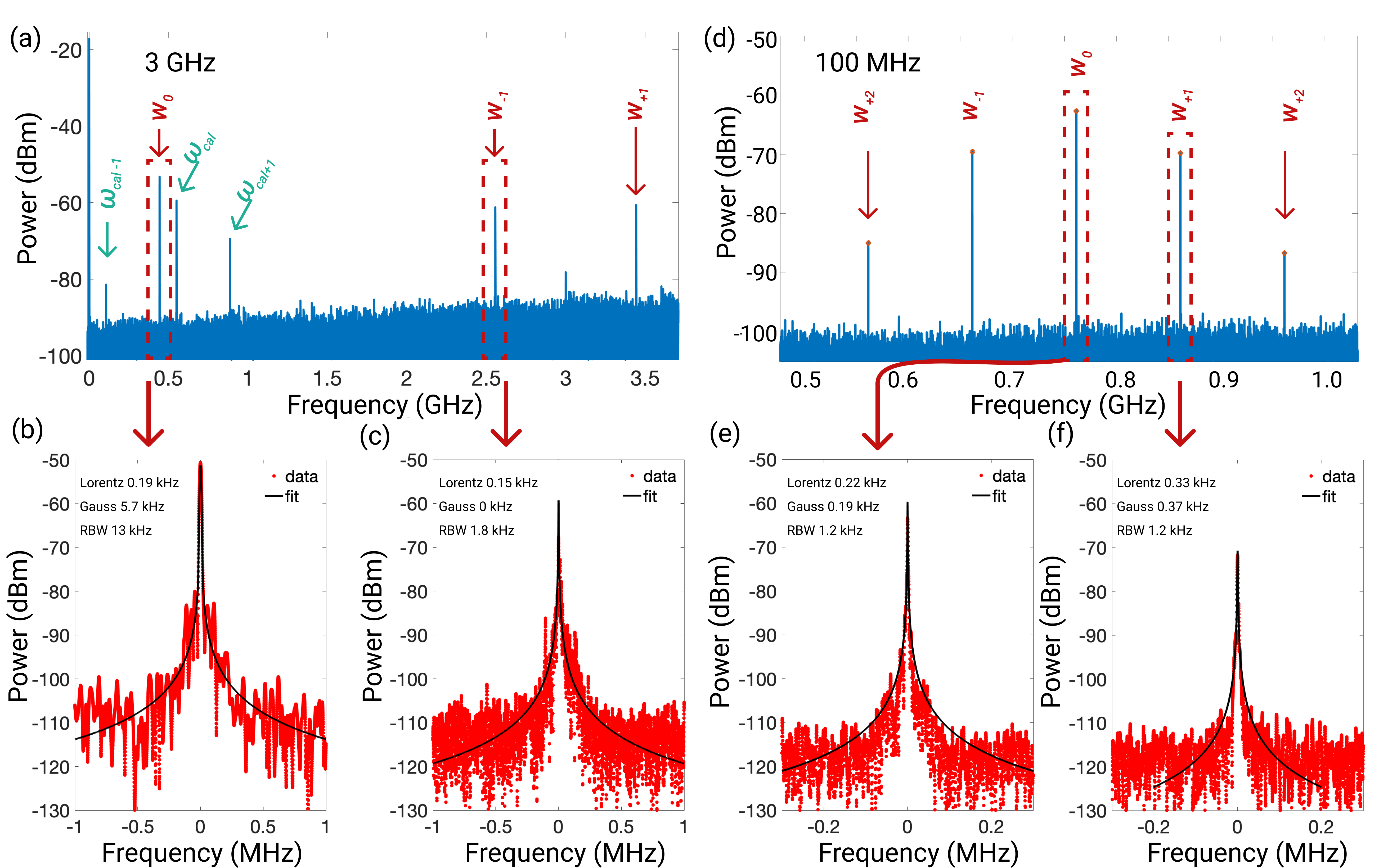}
\caption{The analysis of the teeth linewidth of the GS combs in the SIL regime for  high (3 GHz) and low (100 MHz) modulation frequency. We fit the beatnote signal with Voigt profile. The central line and sideband both have narrow kHz-scale linewidth. }
\label{fig:Ultimate2}
\end{figure*}

\section{SIL impact on the gain switching}

\begin{figure}[hbtp!]
\centering
\includegraphics[width=0.95\linewidth]{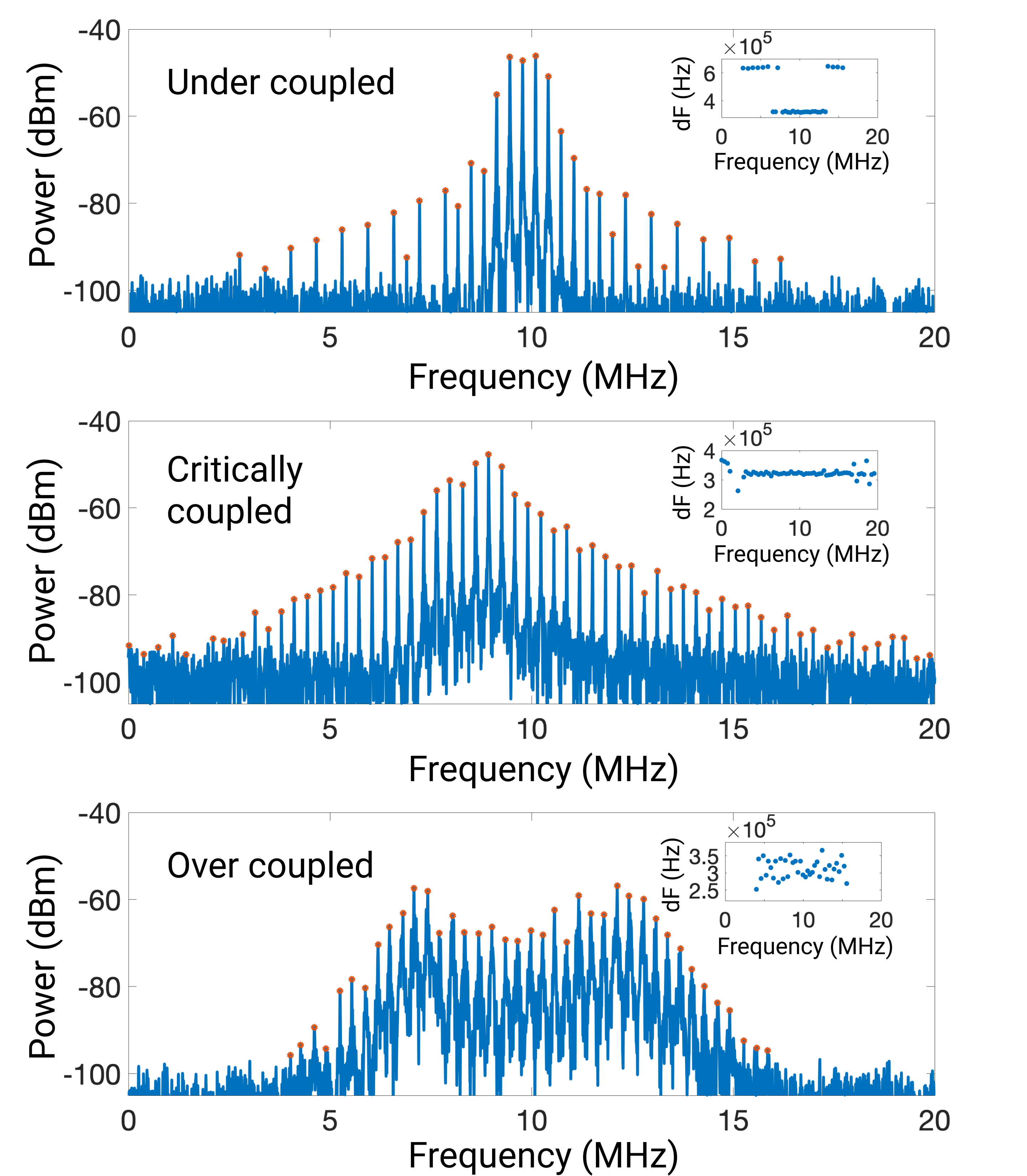}
\caption{  RF spectrum evolution with variation of the coupling efficiency.  }
\label{fig:Coupl}
\end{figure}

Self-injection locking of the GS laser diode to a high-Q WGM drastically changes its spectrum.

First, we compared complete spectra of the locked and free-running gain-switched lasers.
In Fig. \ref{fig:OnOff} the spectra of the modulated GS laser without SIL in the top panel and with it in the bottom panel are presented. SIL was switched off by receding the microresonator from the coupler. The modulation frequency was equal to 100 MHz and the voltage amplitude was 8 V. One may see that while the width of the spectrum does not change a lot  (a broadening of about 30$\%$ can be observed in the SIL regime), a drastic enhancement of the comb contrast by $\approx$20 dBm is clearly visible due to the significant narrowing of the comb teeth. The same effects were observed for the whole range of the studied modulation frequencies. Parasitic lines below 1 GHz $\omega_{\rm gen}, 2\omega_{\rm gen}, 3\omega_{\rm gen}$ and $\omega_{\rm cal}$ can be figured out from the bottom graph. The $\omega_{\rm cal}$ is the Koheras calibration line appearing at 550 MHz and $\omega_{\rm gen}$ is the line corresponding to the modulation frequency.

Then, the spectral properties of the central line and sidebands in case of either low and high modulation frequencies were accurately measured (see Fig. \ref{fig:Ultimate2}). We analyzed the spectral properties of the GS SIL combs for significantly different cases exploring beatnotes of the central line (Fig. \ref{fig:Ultimate2}a) and sideband (Fig. \ref{fig:Ultimate2}b, c) from the 3-GHz frequency comb (modulation voltage of 1.8 V) and central line and sideband from the 100-MHz comb (modulation voltage of 1.4 V, see Fig. \ref{fig:Ultimate2} d, e, f). To analyze the beatnote data we utilized an approximation with Voigt profile (black curves in the panels). The Faddeeva error function was used to calculate a Voigt approximation \cite{ABRAROV20111894}. Voigt profile allows to estimate the contribution of the white frequency noise impact (Lorentzian linewidth) and the flicker frequency noise impact (Gaussian linewidth) \cite{galiev2018spectrum}). The first one is less than 0.5 kHz, the second one is less than 6 kHz.
The obtained results allow to declare the possibility of the generation of the high-contrast electrically-tuned frequency combs with the spectral components of the kHz width. The self-injection locking provides spectral purification of the gain-switched combs for a wide range of the interline frequencies.

We also varied the coupling of the microresonator to the optical pump from the undercoupled regime (losses for coupling less than internal losses) through critical coupling (losses for coupling are equal to internal losses) to overcoupled regime (losses for coupling are greater than internal losses) by changing the gap between the coupling prism and microresonator in approximately $0...200$ nm range (see Fig. \ref{fig:Coupl}). The modulation frequency was equal to 320 kHz and the voltage amplitude was 1.4 V. The width of the full spectrum is almost the same in all three cases. The frequency distance between the lines in overcoupled regime is a lot less stable, see the insets. The noise floor in overcoupled regime is higher and the lines are less narrow. Thus the loading of the microresonator determines the linewidths of the comb teeth and its stability. Note, that in undercoupled regime the remote lines separation is doubled.

\section{Electrical tuning of the gain-switched combs}

\begin{figure*}[hbtp!]
\centering
\includegraphics[width=\linewidth]{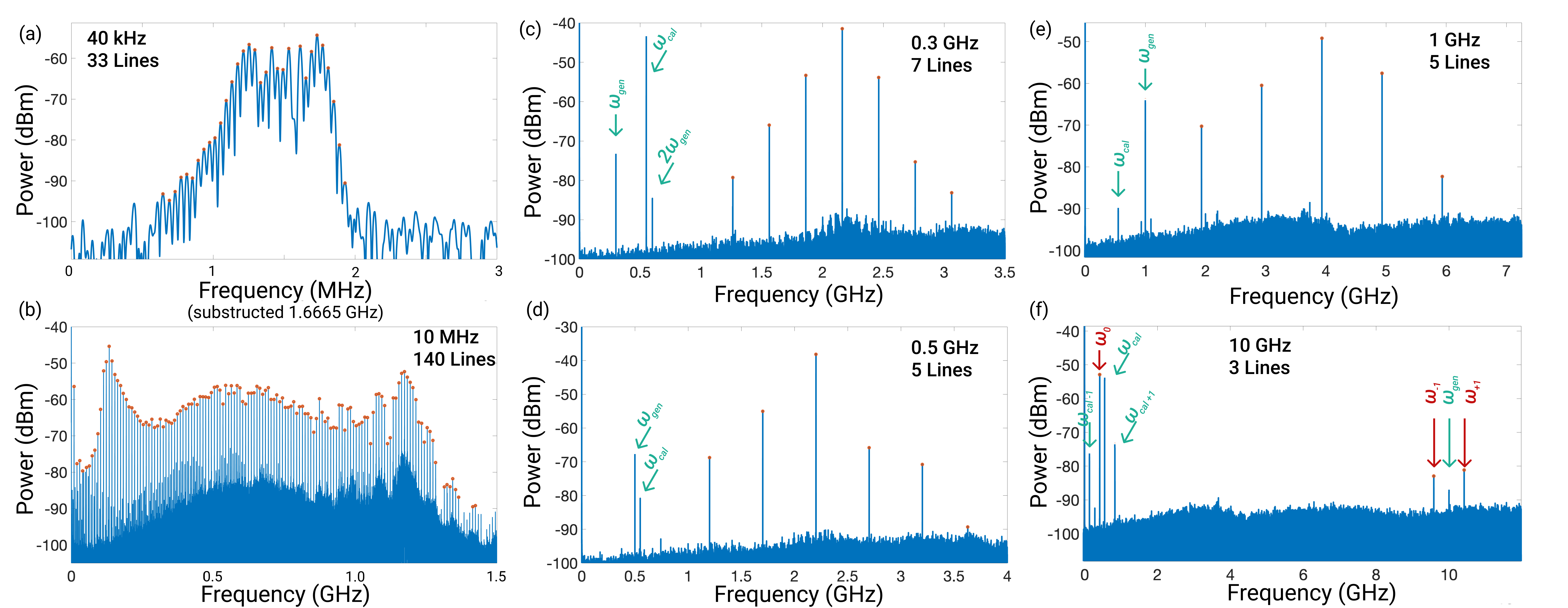}
\caption{The frequency combs observed in SIL regime with gain-switched laser. Spectra were obtained by means of the heterodyne method. Comb teeth are marked with red circles and red arrows, and green arrows correspond to the interference lines. (a) The 40 kHz spaced comb. The spacing between adjacent lines is presented on the inset. (b) The 10 MHz spaced comb with width wider than 1 GHz. (c). The 0.3 GHz spaced comb. (d) The 0.5 GHz spaced comb. (e) The 1 GHz spaced comb. (f) The 10 GHz spaced comb. 
}
\label{fig:SIL}
\end{figure*}

\begin{figure}[hbtp!]
\centering
\includegraphics[width=\linewidth]{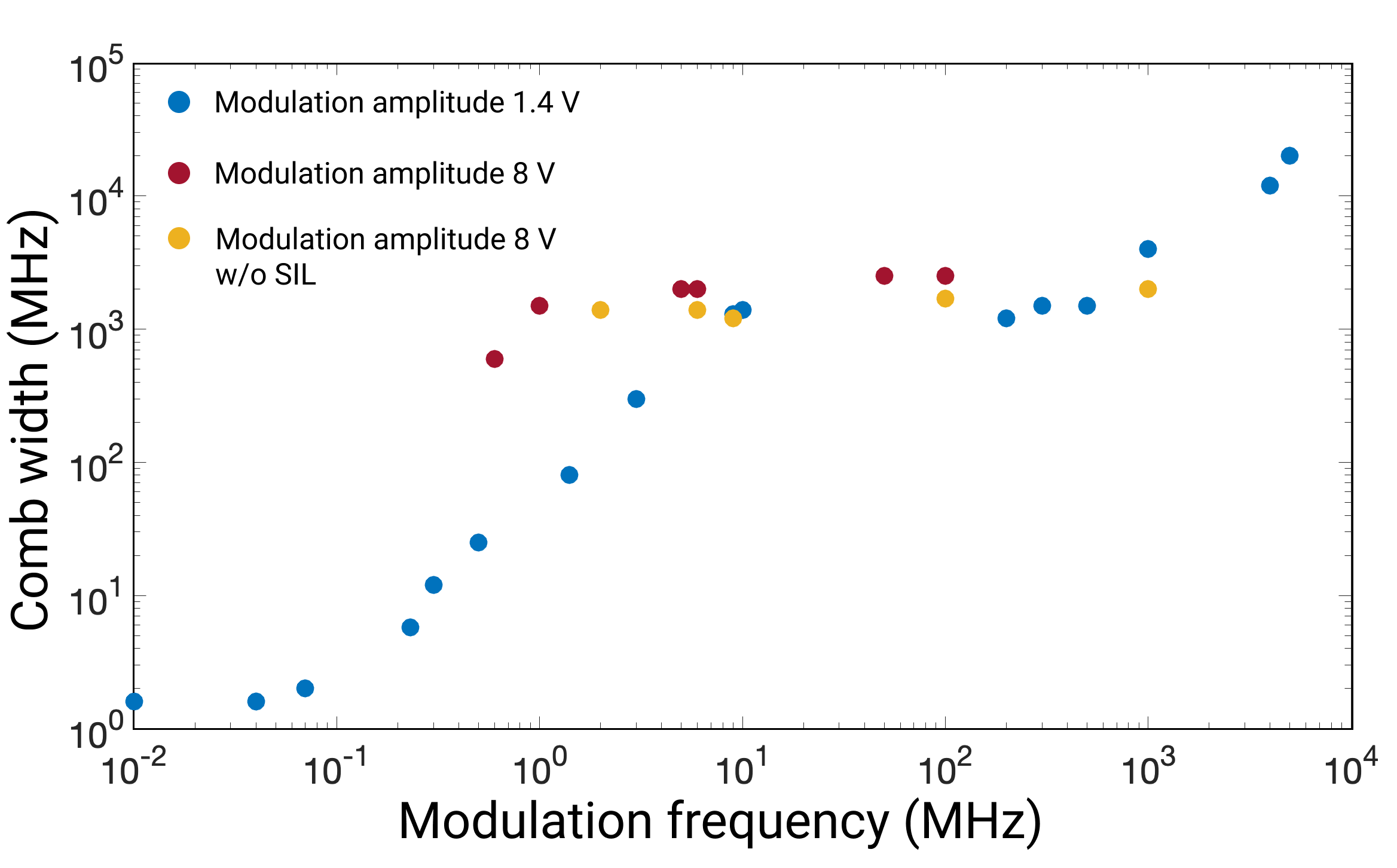}
\caption{ The dependence of the GS comb full width at $-45$ dBm level on the switching frequency in SIL regime. The GS spectrum width without SIL is presented with yellow dots.
}
\label{fig:CombWidth}
\end{figure}

\begin{figure}[hbtp!]
\centering
\includegraphics[width=0.9\linewidth]{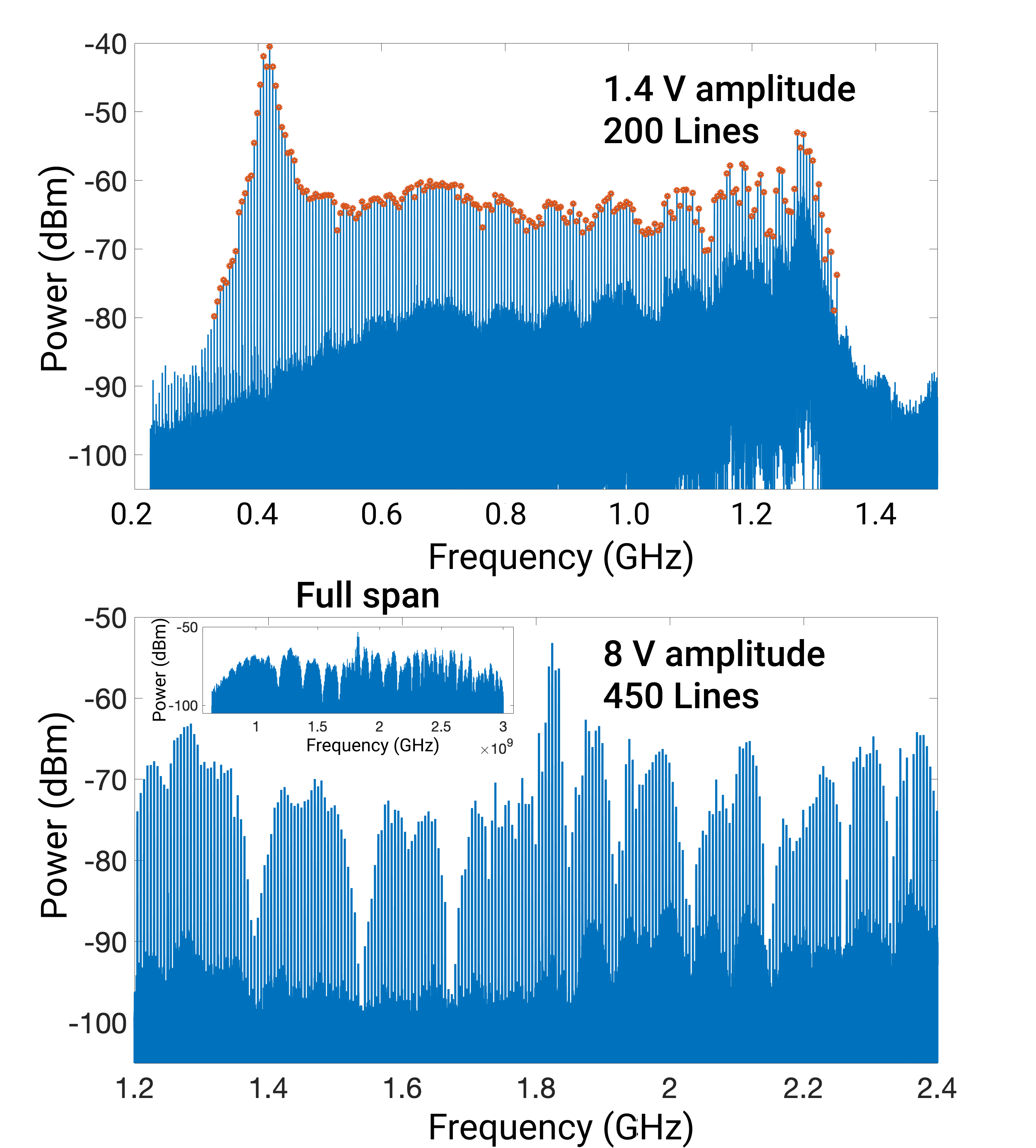}
\caption{ The spectrum broadening with the modulation voltage increase in SIL regime. Upper panel corresponds to the voltage of 1.4 V, bottom panel - to 8.0 V at 5 MHz frequency.  }
\label{fig:Electrically_tuned}
\end{figure}

We studied the characteristics of the generated combs in SIL regime for different modulation parameters. For detailed analysis we varied the modulation frequency and modulation voltage in the wide range and observed frequency combs of different types with the comb line spacing from 10 kHz to 10 GHz (see Fig. \ref{fig:SIL}). 
First, we investigated the evolution of the generated combs with the growth of the modulation frequency. 
 In the panel (a) of Fig. \ref{fig:SIL} a frequency comb with 40 kHz interval between the lines consist of 33 resolvable lines is presented. The modulation amplitude was equal to 1.4 V. The lines at such low repetition rate are indistinguishable for the free-running GS laser and become pronounced due to the linewidth narrowing and frequency stabilization in the SIL regime. In the panel (b) of Fig. \ref{fig:SIL} a frequency comb with 10 MHz interval between the lines and broad spectrum of 1.4 GHz is presented. The modulation amplitude on the generator output was equal to 8 V which led to the spectrum broadening. In the panel (c) and (d) the spectra of the combs with at modulation voltage was 1.6 V and line spacing of 0.3 and 0.5 GHz are presented. The spectra have triangle shape and almost the same width. In the panel (e) a frequency comb with 1 GHz interval between the lines and 4 GHz width is presented. The modulation amplitude was limited by the generator and equal to 1.6 V. Parasitic lines are marked green. They are the Koheras calibration line $\omega_{\rm cal}$ appearing at 550 MHz and the line corresponding to the modulation frequency $\omega_{\rm gen}$. In the panel (f) a frequency comb with 10 GHz interval between the lines is presented. Here the heterodyne is near the central comb line (shown with red arrow $\omega_0$), so both its wings are to the right of it (shown with red arrows $\omega_{\pm 1}$). The modulation amplitude was equal to 1.6 V. The lines, corresponding to $\omega_{cal}$ and its interference with the central line (shown with the green arrows $\omega_{cal \pm 1}$) together with $\omega_{gen}$ appear in that case.

The frequency interval between the lines of the comb can be seamlessly tuned by the GS frequency. However, one should keep in mind that the SIL of GS laser diode is affected by the amplitude of the applied signal. First, it is contributed to biasing the diode that changes its unlocked frequency, second, overdrive may suppress the carrier and a sideband became locked instead. In our case, the bias tee circuit input impedance was frequency dependent and unmatched at high frequencies, so changing the GS frequency we inevitably changed its amplitude too. A frequency-dependent DC current correction and perfect impedance matching leads to ultra-wideband continuous tuning of the SIL GS comb. We managed to keep SIL regime continuously changing the frequency for hundreds of MHz adjusting the GS voltage within 25\%.
 
Thus, we demonstrated that the frequency interval between the lines can be continuously tuned in a vast range if the impedance of the RF circuit is taken into account. The tuning may be either continuous or discrete: we tested tuning steps from 1 MHz up to 100 MHz. On the other hand, the central frequency cannot be tuned in a range larger than several loaded microresonator linewidths ($\approx 1$ MHz in our case, see Fig. \ref{fig:CombsSIL} (e)) without special tricks like simultaneous tuning of the WGM microresonator temperature \cite{Liang2015compact} and laser diode supply current or deformation of the microresonator \cite{liu2020monolithic}. 

In Fig. \ref{fig:CombWidth} the dependence of the comb full width at $-45$ dBm level from a switching frequency is presented. We investigated this dependence for a wide range of the switching frequencies in SIL regime. We also implemented different voltage amplitude (1.4 V marked with red circles and 8 V marked with blue ones in Fig. \ref{fig:CombWidth}) and found the same behavior. The width of the comb was monotonously increased up to the width of 3 GHz at modulation frequency of 10 MHz both due to the interline distance and line number increase. At this level a comb width reached the plateau and we observed the decrease of line number with the growth of the modulation frequency. This plateau is determined by the laser properties and is independent from SIL, because we observed the same spectral width of the generated comb in the unlocked regime (see yellow points in Fig. \ref{fig:CombWidth}). At modulation frequencies of the GHz scale the comb full width again increases with modulation frequency.

Second, we studied the evolution of the generated combs with the variation of the modulation depth and revealed that the amplitude of the modulation voltage is one of the key parameters of the scheme allowing to manage the frequency combs characteristics. In Fig. \ref{fig:Electrically_tuned} the 5-MHz GS combs for the modulation voltage of 1.4 V in the top panel and 8 V amplitude in the bottom panel are presented. For 8 V switching voltage the laser current monitored from the laser diode pins was reaching zero value during oscillation. In that case the width of the spectrum is almost 3 times wider for the same comb line spacing. The modulation voltage can be selected in a manner to converse all the central line into sidebands.

\begin{figure}[hbtp!]
\centering
\includegraphics[width=\linewidth]{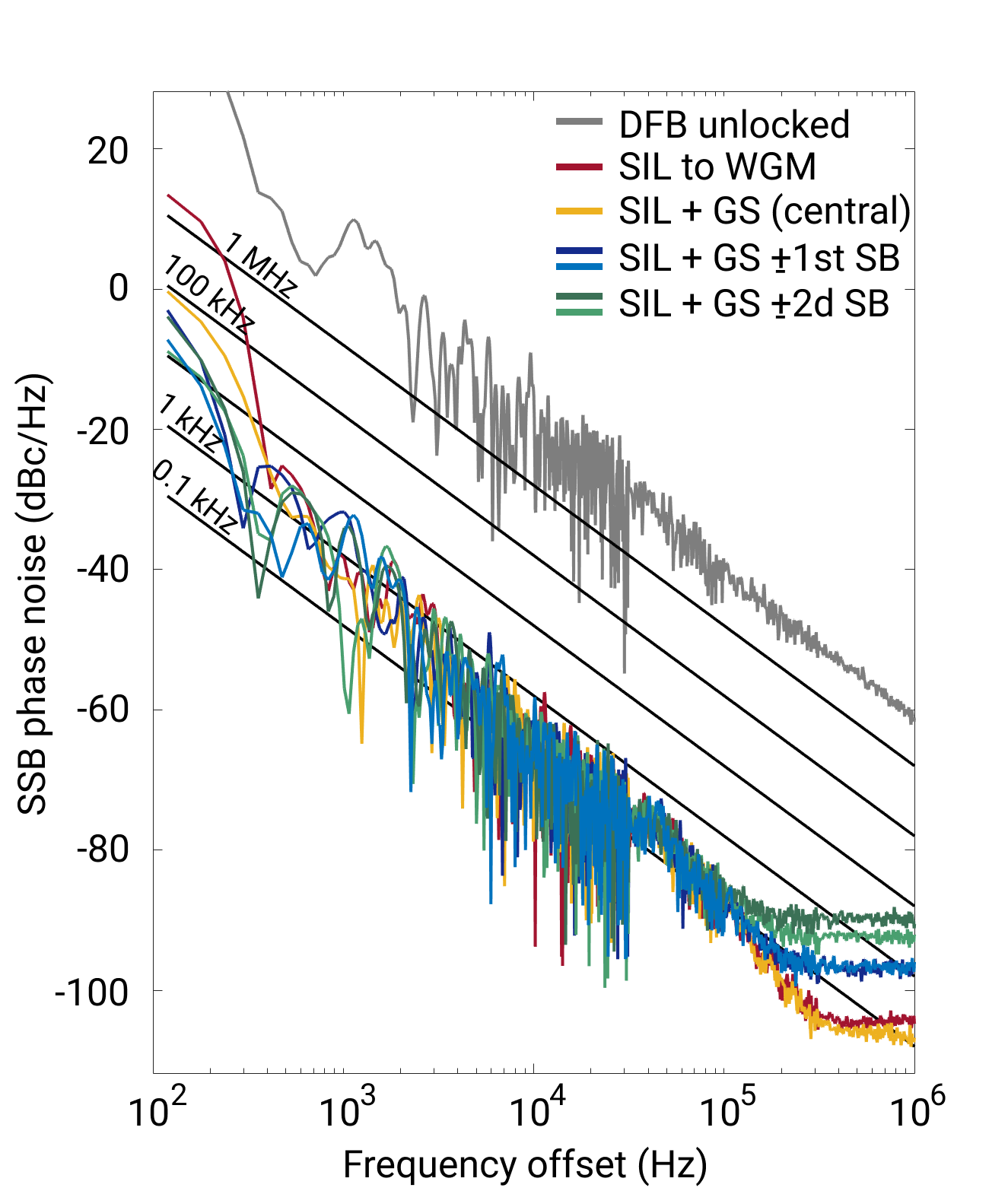}
\caption{Single sideband spectral density of the phase noise of the beat signal of the reference laser and the laser under study in different regimes. Grey line is an unlocked DFB laser. Red line is a SIL laser without modulation applied. The rest is the center line and two sidebands on each side of SIL GS laser. Black lines are the asymptotes corresponds to linewidth of 0.1, 1, 10, 100 kHz and 1 MHz. It can be seen that for frequencies above $10^3$, the values are below 1 kHz asymptote. SIL itself stabilizes at short averaging times and above $10^3$ the fluctuations due to temperature instability are dominant. The lower the line power, the higher the plateau for frequencies closer to MHz. }
\label{fig:phaseNoise}
\end{figure}

The stability of the SIL gain-switched comb was characterized by measurement of the beatnote phase noise (see Fig. \ref{fig:phaseNoise}). The single-sideband phase noise spectral density of the beatnote signal of the laser under consideration and reference laser (Koheras Adjustik) was measured. We compared phase noise level for the unlocked laser, SIL laser without modulation and 100 MHz gain-switched SIL comb central line, first and second sidebands. Until frequency offset $10^3$ Hz, fluctuations of the temperature of the microresonator are the dominant destabilizing factor. Then the phase noise values in locked cases are well below 1 kHz asymptote. We compared the phase noise with the phase noise of two reference lasers and evaluated that in the region from $10^3$ Hz to $10^5$ Hz the phase noise reached the level of the reference laser. At the frequency offset of $10^6$ Hz phase noise value reaches a plateau determined by intensity noise. The sidebands reached the plateau at higher level due to lower peak contrast. This experiment demonstrates outstanding stability of the SIL gain-switched frequency comb.

At this point we demonstrated conceptually the frequency distillation of the gain-switched comb in the self-injection locking regime. The frequency interval between comb lines can be varied in extremely large range from tens of kHz up to tens of GHz. Ordinary equipment (cables, connectors) used in RF scheme were not designed to perform microwave measurements. Thus at GHz frequencies the losses in cable traces became extremely significant, and optimization of the RF path may lead to even better performance.

\section {Conclusion}
GS laser optical frequency combs were observed experimentally with an ordinary continuous-wave DFB laser locked to a high-Q WGM microresonator. We revealed an immense flexibility of the switching rate (from 10 kHz to 10 GHz in our case).  We revealed that SIL leads to a frequency distillation of the comb teeth up to kHz scale and to significant increase of the comb contrast (by more than 20 dBm in our case). The linewidth of the central line and the comb teeth were measured as narrow as in non-switched SIL case. We disclosed that the loaded Q-factor of the microresonator defines the linewidth narrowing and its stabilization as in a plain SIL. The comb line spacing value is determined by the modulation frequency from a generator and can be continuously tuned. On the other hand the full width of the comb can be controlled by adjusting the amplitude of the modulation voltage. That results may be undoubtedly useful for coherent communications, spectroscopy, quantum key distribution or in coherent LIDARs.

\section*{Acknowledgements}
The work was supported by the Russian Science Foundation (project 20-12-00344).

\newpage
\eject

\bibliography{Textbib}

\end{document}